\documentstyle[amssymb,draft,12pt,epsf,palatino]{nature-pvd}

\def\simlt{\mathrel{\hbox{\rlap{\hbox{\lower4pt\hbox{$\sim$}}}\hbox{$<$}}}}
\def\simgt{\mathrel{\hbox{\rlap{\hbox{\lower4pt\hbox{$\sim$}}}\hbox{$>$}}}}

\def\ale{\mathrel{\hbox{\rlap{\hbox{\lower4pt\hbox{$\sim$}}}\hbox{$<$}}}}
\def\age{\mathrel{\hbox{\rlap{\hbox{\lower4pt\hbox{$\sim$}}}\hbox{$>$}}}}

\def\kms{km\,s$^{-1}$}
\def\msun{M$_{\odot}$}
\def\g1256{1255--0}

\def\spose#1{\hbox to 0pt{#1\hss}}
\newcommand\lsim{\mathrel{\spose{\lower 3pt\hbox{$\mathchar''218$}}
     \raise 2.0pt\hbox{$\mathchar''13C$}}}
\newcommand\gsim{\mathrel{\spose{\lower 3pt\hbox{$\mathchar''218$}}
     \raise 2.0pt\hbox{$\mathchar''13E$}}}

\begin{document}

\title{\LARGE \textsf{A high stellar velocity dispersion for a
compact\vspace{0.15cm}\\massive galaxy at z\,=\,2.2}}

\author{\textsf{Pieter G.\ van Dokkum}\affiliation[1]
  {\textsf{Astronomy Department, Yale University, 260 Whitney Ave, New Haven,
   CT 06511, USA}},
   \textsf{Mariska Kriek}\affiliation[2]
   {\textsf{Department of Astrophysical Sciences, Princeton University,
 Princeton, NJ 08544, USA}},
   {\textsf{\&} \textsf{Marijn Franx}}\affiliation[3]
   {\textsf{Leiden Observatory, Leiden University, NL-2300 RA Leiden,
The Netherlands}}
}
\date{\today}{}
\headertitle{Velocity Dispersion of a Compact Galaxy}
\mainauthor{van Dokkum et al.}

\summary{Recent studies have found that the oldest and most luminous
galaxies in the early Universe are surprisingly compact,
\cite{trujillo:06,toft:07,vd:08,cimatti:08,wel:08,franx:08,dam:09}
having stellar masses
similar to present-day elliptical galaxies but much smaller
sizes.
This finding has attracted considerable
attention\cite{naab:07,fan:08,bezanson:09,naab:09,wel:09,hopkins:09}
as it suggests that massive galaxies have grown
by a factor of $\sim$\,five in size over the past ten
billion years. A key test of these results is a determination of the
stellar kinematics of one of the compact galaxies: if the
sizes of these objects are as extreme as has been claimed,
their stars are expected to have much higher velocities
than those in present-day galaxies of the same mass.
Here we report a measurement of the stellar velocity dispersion
of a massive compact
galaxy at redshift ${\bf z=2.186}$,
corresponding to a look-back time of 10.7 billion years.
The velocity dispersion is very high at
${\bf 510^{+165}_{-95}}$\,\kms,  consistent with the mass and compactness
of the galaxy inferred from photometric data and indicating significant
recent structural and dynamical evolution of massive galaxies.
The uncertainty in the dispersion was determined from
simulations which include the effects of noise and
template mismatch.
However, we caution that some subtle systematic effect may influence the
analysis given the low signal-to-noise ratio of our spectrum.}

\maketitle


We observed the galaxy, dubbed \g1256, with the Gemini Near-Infrared
Spectrograph (GNIRS) on the Gemini South telescope for a total of 29\,hrs.
The de-redshifted spectrum is shown in Fig.\ 1a.
A detailed description of the observations and reduction,
as well as an analysis of the continuum
emission and detected (weak) emission lines,
is presented in a companion paper.\cite{kriek:09}
In the companion paper we derive a stellar mass of
$\approx 2.0 \times 10^{11}$\,\msun\ for a Kroupa initial mass
function, by
fitting stellar population synthesis models to the
broad band photometry and the GNIRS spectrum.
The effective radius of \g1256\ $r_e = 0.78 \pm 0.17$\,kpc, as
previously measured\cite{vd:08} from
deep Hubble Space Telescope (HST) NICMOS2 observations. 
The galaxy was selected from a
well-studied\cite{kriek:06,kriek:08,vd:08} sample of nine
spectroscopically-confirmed galaxies with evolved stellar
populations at
$z\sim 2.3$, and its properties are similar to those of other galaxies
in this sample. The median stellar mass of the nine objects is
$1.7 \times 10^{11}$\,\msun\ and their median effective radius
$r_e=0.9$\,kpc\cite{vd:08}, a factor of $\sim 5$ smaller than galaxies with
similar masses at $z=0$.
The number density of these massive compact galaxies is substantial,
about the same as that of galaxies in the nearby Universe that are
a factor of 2--3 more massive.\cite{bezanson:09}

In the present study we use the deep Gemini spectrum
to measure the stellar velocity dispersion of the galaxy, by employing
standard techniques for measuring the broadening of the absorption
lines.\cite{fih:89,vd:03} Our methodology is explained in detail
in the Supplementary Information. Briefly,
smoothed model spectra were fitted to the data
in real space, taking observational errors into account
and ignoring data with the largest uncertainties.
The uncertainty in the dispersion
was determined from Monte Carlo simulations of many different
combinations of assumed velocity dispersions and empirical
realizations of the noise. Systematic uncertainties
were assessed by varying
the templates (also allowing for multiple components),
the masking and weighting, and the continuum filtering, and
are typically much smaller than the random uncertainty.
We note that the spectrum is available in electronic form
as a Supplementary Dataset.

We derive a velocity dispersion $\sigma = 510^{+165}_{-95}$\,\kms\
for the galaxy, which is very
high when compared to typical early-type
galaxies in the nearby Universe. 
Although not statistically
significant, it is striking that
the best-fit value exceeds the measured dispersions of all
individual galaxies in the Sloan Digital Sky Survey
(SDSS).\cite{bernardi:06,bernardi:08}
In the SDSS
a significant fraction of galaxies with velocity dispersions in
excess of 350\,\kms\ are superpositions, which are easily identified
with HST imaging.\cite{bernardi:08} As shown in Fig.\ 1b-d,
\g1256\ is a single, nearly round object with an effective radius of
$\approx 0.1$ arcsec in HST images. The dispersion is
also a factor of $\sim 2$
higher than a previous measurement\cite{cenarro:09} from
a stacked spectrum of 13 galaxies at
$\langle z \rangle = 1.6$. A direct comparison is difficult given the
uncertainties associated with stacking individual spectra, but we note
that the median stellar mass of the 13 galaxies is a factor of $\sim 3$ smaller
than that of \g1256\ and the median effective radius is a factor of 1.5
larger. The expected dispersion of these
$\langle z \rangle = 1.6$ galaxies is therefore a factor of $\sim 2$ lower
than that of \g1256, and the two results are consistent.

The high dispersion of \g1256\ confirms that the galaxy is
very massive despite its diminutive size. The relation between mass,
velocity dispersion, and size can be expressed as $M_{\rm dyn} =C \sigma^2
r_e$, with $C$ a constant that depends on the structure of the galaxy
and other parameters. Using $\log C = 5.87$, which is the
value that gives $M_{\rm dyn} \approx M_{\rm star}$ for galaxies in the
SDSS\cite{franx:08}, we find
$M_{\rm dyn} = 1.5^{+1.2}_{-0.5} \times
10^{11}$\,\msun.
For $\log C=6.07$, the value derived from kinematic
data of present-day early-type galaxies,\cite{vd:03} the dynamical mass
is $2.4^{+1.9}_{-0.8} \times 10^{11}$\,\msun. Both estimates
are in excellent agreement with the stellar mass (Fig.\ 2a).
Put differently, the high dispersion
that we measure was {\em expected} (and in fact
predicted\cite{vd:08,bezanson:09})
given our extreme size and stellar mass measurements.
Quantitatively, the expected dispersion assuming $M_{\rm dyn} =
M_{\rm star}$ and $5.87 \leq \log C \leq 6.07$
is in the range 470\,\kms\ -- 590\,\kms.

At the same time, the high dispersion confirms and extends the notion
that quiescent galaxies at high
redshift are structurally and dynamically very different from galaxies in
the present-day Universe.
Figures 2b-d show
where \g1256\ falls with respect to
the relations between velocity dispersion, size,
and dynamical mass defined by SDSS galaxies.
The galaxy is
offset from the local relations,
consistent with previous
studies which were based on stellar masses derived
from photometric data.\cite{vd:08,franx:08} At fixed dynamical mass
the dispersion is higher by a
factor of $\sim 2.5$ and the effective radius is smaller
by a factor of $\sim 6$. The most dramatic offset is in the
$\log \sigma - \log r_e$ plane (Fig.\ 2b). These two quantities
are measured directly and independently, and (to first order)
do not depend on stellar populations.

The extreme compactness of massive high redshift galaxies
is qualitatively consistent with
models in which the central parts of massive galaxies form early in
highly dissipative processes\cite{dekel:09}, although it remains
to be seen whether such models can produce objects with
the size and velocity dispersion of \g1256. In particular, it
may be difficult to funnel gas clumps into an extremely
compact configuration without forming stars at larger radii.
Regardless of the details of the model,
in its star-forming phase at $z\gtrsim 3.5$ the galaxy
likely had a very compact
molecular gas distribution with a rotation velocity of
$\sim 700$\,\kms.  The median rotation
velocity of CO in submm galaxies at $z=2-3.5$
has been found\cite{greve:05} to be $\approx 470$\,\kms\
(assuming $V_{\rm rot} = 0.6 \times {\rm FWHM}$) --- a high value
by most standards, but still somewhat lower than what we expect for the
progenitors of galaxies such as \g1256.
There is not yet much
information on the gas dynamics of massive galaxies at redshifts
$z> 3.5$. The $z=6.4$ quasar SDSS\,J1148+5251 has a relatively
small CO linewidth of $V_{\rm rot} \approx 170$\,\kms,\cite{bertoldi:03}
but it may be that
quasars are biased low because their gas disks are preferentially
seen face-on.\cite{narayanan:08}
It is obviously not clear
whether the gas was ever in a regular disk;
it would be interesting to determine whether  \g1256\ shows
rotation, but that requires imaging (or spectroscopy) of
higher spatial resolution than is currently available.

A problem that is perhaps even more vexing than the origin of
galaxies such as \g1256\ is their subsequent
evolution onto the local relations between size, velocity dispersion,
and mass. The simplest explanation is that the mass and/or size
measurements of the compact galaxies are incorrect\cite{vd:08,hopkins:09},
but this is
difficult to maintain given the dynamical measurement presented here.
We are left with the conclusion that
very significant structural and dynamical changes are required
to bring massive, quiescent high redshift
galaxies to the local relations.
This cannot easily be achieved through star formation as the compact
high redshift galaxies already appear to have stopped forming
new stars, consistent with the old ages 
inferred for the stars in today's most massive galaxies.
Among the models that have been
proposed\cite{naab:07,fan:08,bezanson:09,naab:09,wel:09,hopkins:09}
minor mergers may be the most effective single
mechanism, as simple virial arguments suggest that the
velocity dispersion changes by a factor of $f_r^{-1/4}$ for
a factor of $f_r$ change in radius.\cite{bezanson:09,naab:09} However,
it is an open question whether mergers alone can ``puff up''
galaxies by the required amount, as
the precise effect  depends on the accretion rate,
the masses, orbits, and gas content
of accreted galaxies, angular momentum transfer between stars
and dark matter, and on possible evolution in the
contribution of dark matter to the measured kinematics.\cite{boylan:06}
Finally, we note that evolution in the velocity dispersion
of galaxies would trivially imply  evolution in the
black hole mass -- $\sigma$ relation,\cite{ferrarese:00,gebhardt:00}
such that black hole masses
are lower at fixed $\sigma$ at high redshift.

While confirming that the velocity dispersions of compact galaxies are
high, our measurement is obviously not sufficiently accurate to
properly characterize the evolution of the relations in Fig.\ 2.
A $1\sigma$ error of 25\,\% in the velocity
dispersion implies an error of $56$\,\%
in the dynamical
mass, and further progress requires dispersions with uncertainties
$\lesssim 10$\,\% for much larger samples.
New spectrographs being readied for use on 8m class
telescopes, combined with new wide field imaging
surveys that can provide
sufficiently bright targets, are expected
to revolutionize this field in the next few years.
As indicated here, such observations are crucial
for calibrating stellar masses at high redshift and for
measuring the structural and dynamical evolution of massive
galaxies from the time that their star formation was quenched
to the present.

\bibliographystyle{nature-pap}

\vspace{0.3cm}

\noindent
{\small \bf \textsf{Supplementary Information}}
\small{\textsf{is linked to the online version of the
paper at www.nature.com/nature.}}
\vspace{0.5cm}\\
{\small \bf \textsf{Acknowledgements}}
{\small \textsf{Based on observations obtained at the Gemini Observatory
and with the Hubble Space Telescope.
This work was supported by NASA and the National Science
Foundation.
We thank Ivo Labb\'e, Garth Illingworth, Danilo Marchesini, and
Ryan Quadri for their contributions in the initial stages of this
project.}}
\vspace{0.5cm}\\
{\small \bf \textsf{Author Contributions}}
{\small \textsf{P.v.D.\ wrote the Gemini proposal,
did the observations, measured the velocity dispersion, wrote the
paper and led the interpretation. M.K.\ reduced the Gemini
spectrum, determined the stellar mass, and contributed to the
interpretation. M.F.\ independently measured the velocity dispersion and
contributed to the analysis and interpretation.}}
\vspace{0.5cm}\\
{\small \bf \textsf{Author Information}}
{\small \textsf{Reprints and permissions information is available at
npg.nature.com/reprintsandpermissions.
Correspondence and requests for materials
should be addressed to P.v.D.\ (pieter.vandokkum@yale.edu).}}

\noindent 
\vspace{-1cm}
\noindent
\begin{figure}
\epsfxsize=14.5cm
\epsffile{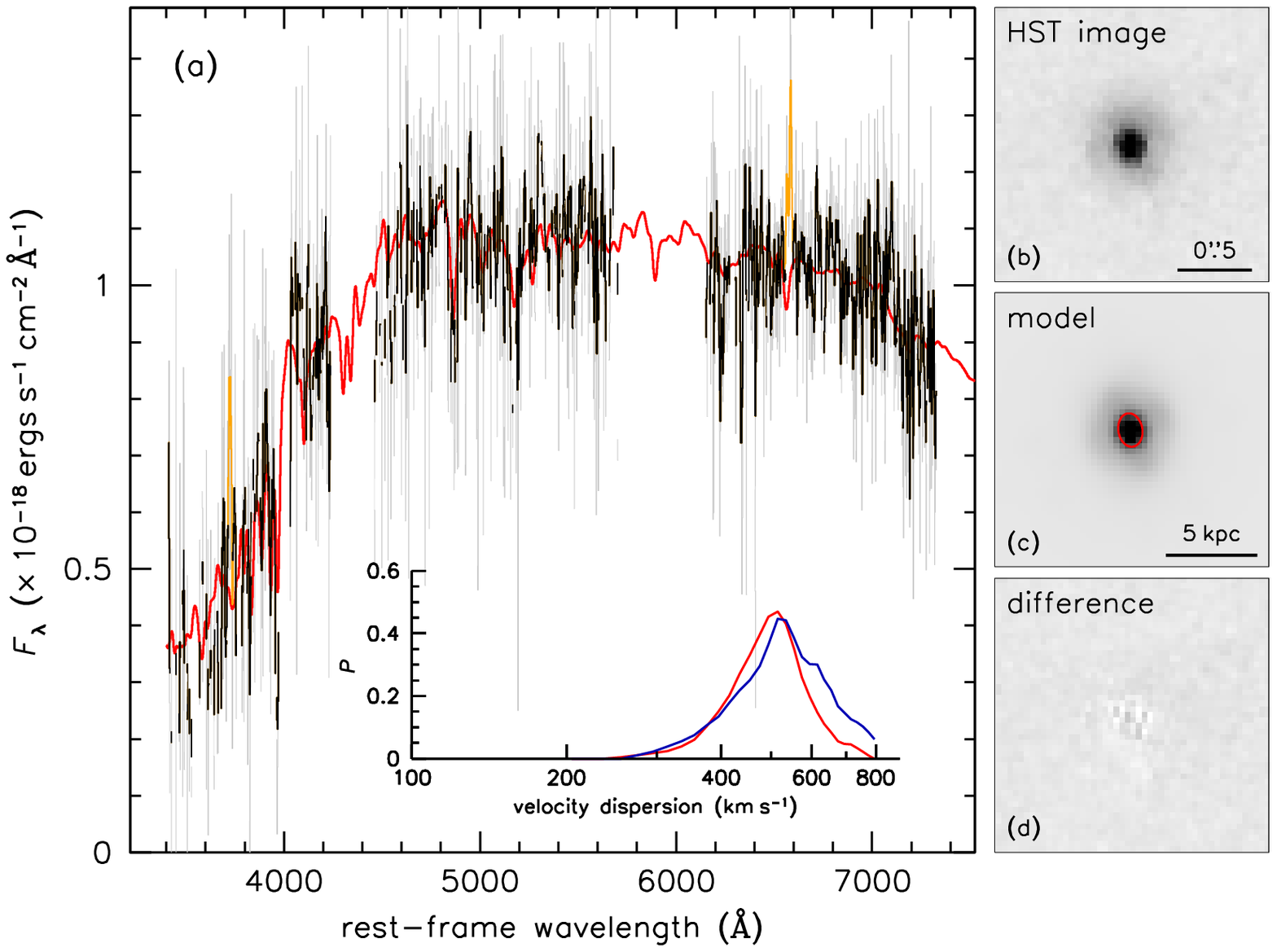}\vspace{0.3cm}\\
{\small \bf {\textsf{Figure 1}} $|$ } 
{\small \bf {\textsf{Spectrum and HST images
of \g1256\ at ${\bf z=2.186}$.}}}
{\small
{\bf a}, Spectrum that was used to measure the velocity dispersion.
Light grey shows
the spectrum at a resolution of 5\,\AA\ ($\approx 100$\,\kms), which
was used for the actual measurement. A smoothed
version of the same data
(using a 25\,\AA\ boxcar filter) is shown in black. Regions around
detected emission lines are shown in orange and were excluded from the
fits. The most prominent
absorption lines are H$\beta$ at $\lambda$4861\,\AA\ and
Mg at $\lambda 5172$\,\AA. The best-fitting stellar population
synthesis model,\cite{kriek:09}
smoothed to the best-fitting velocity dispersion,
is shown in red. The inset shows the results of Monte Carlo
simulations to determine the uncertainty in the best-fitting
velocity dispersion. The curves show how often a dispersion
of $510$\,\kms\ is measured given the true dispersion and noise.
The two curves are for two different methods of simulating
noise: shuffling the residuals of the fit in the wavelength
direction (blue curve), and extracting ``empty'' 1D spectra
from the 2D spectrum (red curve). {\bf b-d}, The HST
NICMOS2 image of the galaxy in the $H_{160}$ filter, the best-fitting
model of the galaxy (with the effective radius indicated in red),
and the residual obtained by subtracting the model from the data.
The galaxy is a single, very compact object with an effective radius
of $0.78$\,kpc. Its coordinates are $\alpha=12^{\rm h}54^{\rm m}59.6^{\rm s}$,
$\delta=+01^{\circ}11^{\rm m}30^{\rm s}$ (J2000), its $K$ band
observed magnitude is 19.26 (Vega) and its $R$ band observed
magnitude is 24.98 (Vega)\cite{kriek:08}. Alternative names that
have been used for this object are 1256-151\cite{kriek:06}
and 1256-0\cite{vd:08,kriek:08}.}
\end{figure}

\bigskip

\noindent
\noindent 
\vspace{-1cm}
\noindent
\begin{figure}
\epsfxsize=14cm
\epsffile{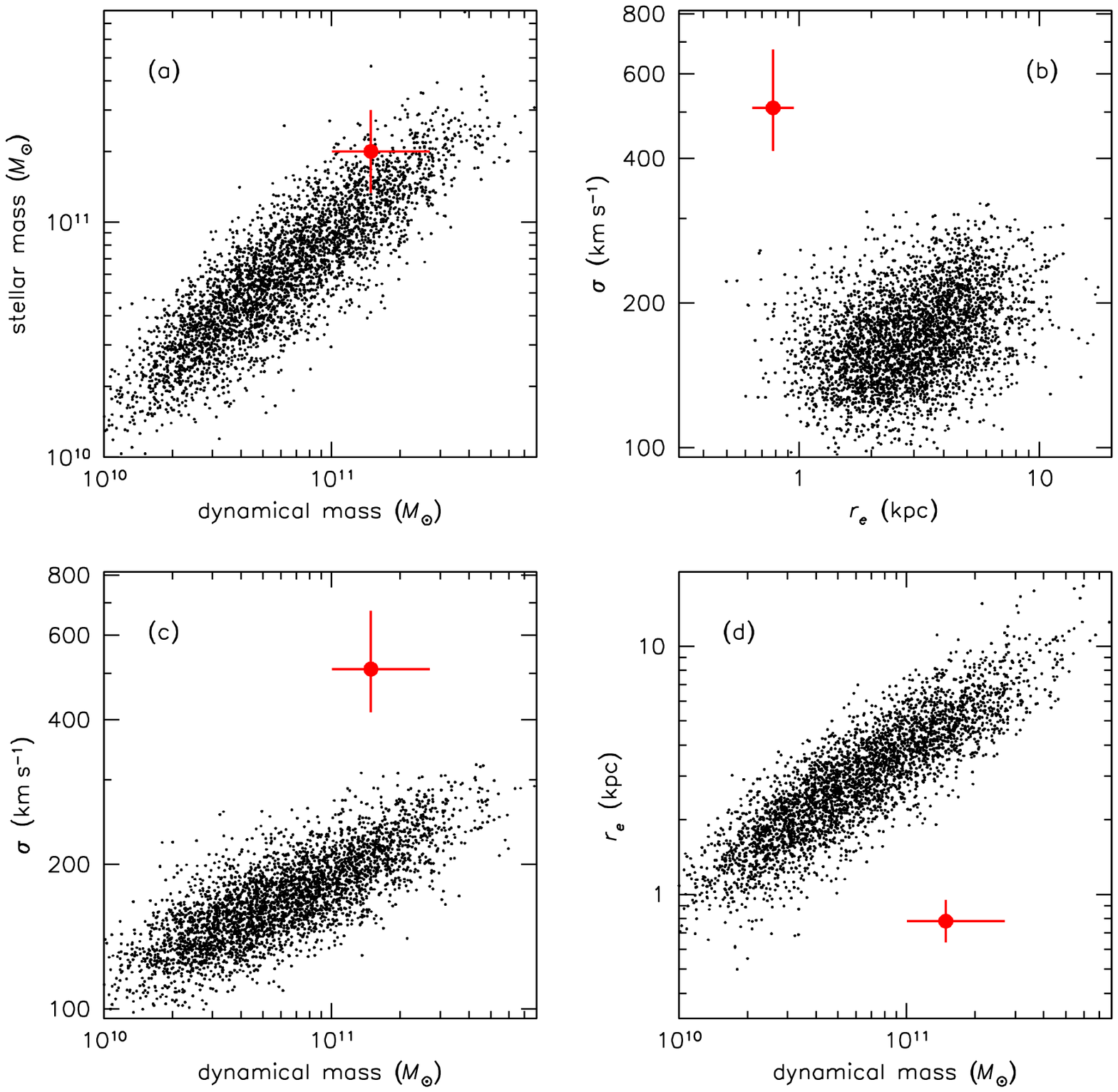}\vspace{0.3cm}\\
{\small \bf \textsf{Figure 2} $|$ }
{\small \bf \textsf{Properties of \g1256\ compared to nearby
galaxies.}}
{\small {\bf a},
Relation between stellar mass and dynamical mass.
Small symbols are galaxies in the SDSS\cite{franx:08} in the redshift range
$0.05-0.07$, and the large red symbol is galaxy \g1256\ at $z=2.186$.
Our definition of dynamical mass,
$\log M_{\rm dyn}
 = 5.87 + 2 \log(\sigma) + \log(r_e)$, 
leads to a one-to-one correspondence
between stellar mass and dynamical mass for SDSS galaxies.
Despite its small size \g1256\ has a very high mass, similar to
elliptical galaxies today. The dynamical mass is consistent
(within $1\sigma$)
with the stellar mass that was estimated\cite{kriek:09} from fitting
stellar population synthesis models to the photometry.
{\bf b-d,} Relations between velocity dispersion,
effective radius, and dynamical mass.
Note that these three panels do not depend on stellar populations
(except indirectly through the fact that
the spectrum and the Hubble Space Telescope image are weighted
by luminosity, not mass). It is clear that the
structure and kinematics of \g1256\ are fundamentally
different from those of nearby galaxies, and significant evolution
is required to bring this object to the local relations.
}
\end{figure}

\newpage

\onecolumn
\noindent
{\bf \textsf{Supplementary Information:
Robustness of the velocity dispersion}}
\vspace{0.2cm}\\
The S/N ratio of the spectrum ranges from 5--8 per resolution element
between the sky lines, which is about a factor of two
lower than what is typically used
for velocity dispersion measurements. We are helped by the unusually
large rest-frame wavelength coverage of the spectrum, the large
value of the dispersion, and the fact that in this case a measurement with a
$1\sigma$ uncertainty of $\sim 20$\,\% is already highly informative.
However, a concern is that systematic effects can dominate
when the S/N is low. Here we describe several tests we performed to
test the robustness of the measurement.

\vspace{0.2cm}
\noindent
{\em Templates:} The default template is the best-fit to the rest-frame
UV -- optical SED. This is a Solar metallicity model with an
age of 2.1 Gyr, an e-folding time $\tau = 0.3$\,Gyr, and extinction
$A_V=0.25$\,mag. We also fitted single stellar populations (SSPs)
with a range of ages and metallicities.
We note that the linestrength is
not fixed but a free parameter in the fit.
The youngest models that were used have
ages of 0.5\,Gyr, despite the fact that such models provide a
poor fit to the overall SED. Composite models were also tried,
comprised of a 0.5\,Gyr component in addition to a maximally-old
population. Finally we fitted several stellar spectra that we
used previously at lower redshifts, even though they span a smaller
wavelength range than the galaxy spectrum.
The velocity dispersion is fairly
insensitive to the choice of template, varying mostly well
within the quoted uncertainty, particularly when the
absorption-line redshift is constrained to be within several hundred
\kms\ of the emission-line redshift. This is illustrated in Supp.\
Fig.\ 1, which shows the reduced $\chi^2$ as a function of velocity
dispersion for the default template and for three examples of
alternative templates.
\vspace{0.2cm}\\
{\em Fitting region:} The S/N ratio of the full spectrum is just
sufficient for a dispersion measurement, and fitting over smaller
wavelength regions gives results that are much less robust.
Most of the signal is contributed by the observed $H$ band: in
the $J$ band the
S/N is lower, and in the $K$ band the absorption lines are weaker.
Fitting to the observed $J + H$ bands gives $\sigma = 480$\,\kms\ and
fitting to the $H+K$ bands gives $\sigma = 590$\,\kms, both with
large uncertainties. We illustrate which wavelength regions
contribute most to our measurement in Supp.\ Fig.\ 2. The only rest-frame
region
that ``prefers'' a significantly smaller dispersion is just blueward
of the 4000\,\AA\ break. This region has relatively low S/N
in our spectrum and it is notoriously sensitive to
template mismatch.
\vspace{0.2cm}\\
{\em Fitting method:} The spectrum was fit in real space, weighting
by the S/N ratio and disregarding data with the highest noise (mostly
the wavelength regions  between the atmospheric windows).
Emission lines were excluded; exluding the H$\beta$
absorption line has no effect on the measured dispersion.
Template spectra were convolved to the instrumental resolution of
$\sigma_{\rm instr} \approx 140$\,\kms\ (taking their intrinsic
resolution into account) and subsequently broadened to a
finely sampled grid of velocity dispersions. Continuum filtering
was done with a 5$^{\rm th}$ order polynominal; we verified that the
results are insensitive to the order of the fit. The redshift and
the normalization of the residual of the continuum fit (i.e., the
linestrength) are free parameters in the fit. Two of us (P.v.D.\ and
M.F.) independently wrote software to fit the spectrum, and our
results are fully consistent.
\vspace{0.2cm}\\
{\em Determination of uncertainty:} The quoted uncertainty was
determined from Monte Carlo simulations. We created a finely
sampled grid of ``true'' dispersions. For each dispersion $\sigma_{\rm true}$
a model spectrum was created. Next, 500 realizations of the model
spectrum were created by adding realistic noise. These 500
noisy spectra were fit with the same procedure as used for the
actual data. The template was allowed to vary in the simulations.
For $\sigma_{\rm true}<510$\,\kms\ the probability
$P(\sigma_{\rm true})$ is then the fraction of simulations that
give a best-fit dispersion in excess of 510\,\kms, and for
$\sigma_{\rm true}>510$\,\kms\ it is the fraction of simulations
that give measured dispersions below 510\,\kms.

The curves in
the inset of Fig.\ 1a show the results of the simulations for
two different ways to create realistic noise spectra. The
simulations shown by the blue curve randomly redistributed the
residuals from the best fit in the wavelength direction. This
approach has the advantages that the noise characteristics of the
data are exactly maintained and that the contribution of template
mismatch to the residuals is included.
We shuffled blocks of 5 pixels (25\,\AA)
rather than individual pixels to retain the effects
of correlated noise. The simulations that gave rise to the red
curve were created by combining random rows of the two-dimensional
spectrum. This approach has the advantage that the wavelength
dependence of the noise is maintained. As shown in Fig.\ 1a
both approaches give very consistent results. The quoted
uncertainty was calculated from the blue curve as it is slightly
wider than the red curve. Both
simulations give a slightly larger uncertainty
than the formal error from the $\chi^2$
distribution (shown in Supp.\ Fig.\ 2). Finally, we note that
the true uncertainty may be (even) larger than the quoted
uncertainty
due to some unknown systematic effect; as is well known,
systematic effects become increasingly important at low
S/N ratios. We
provide the spectrum in electronic form as a
Supplementary Dataset.

\vspace{-1cm}
\noindent
\begin{figure}
\epsfxsize=13cm
\epsffile{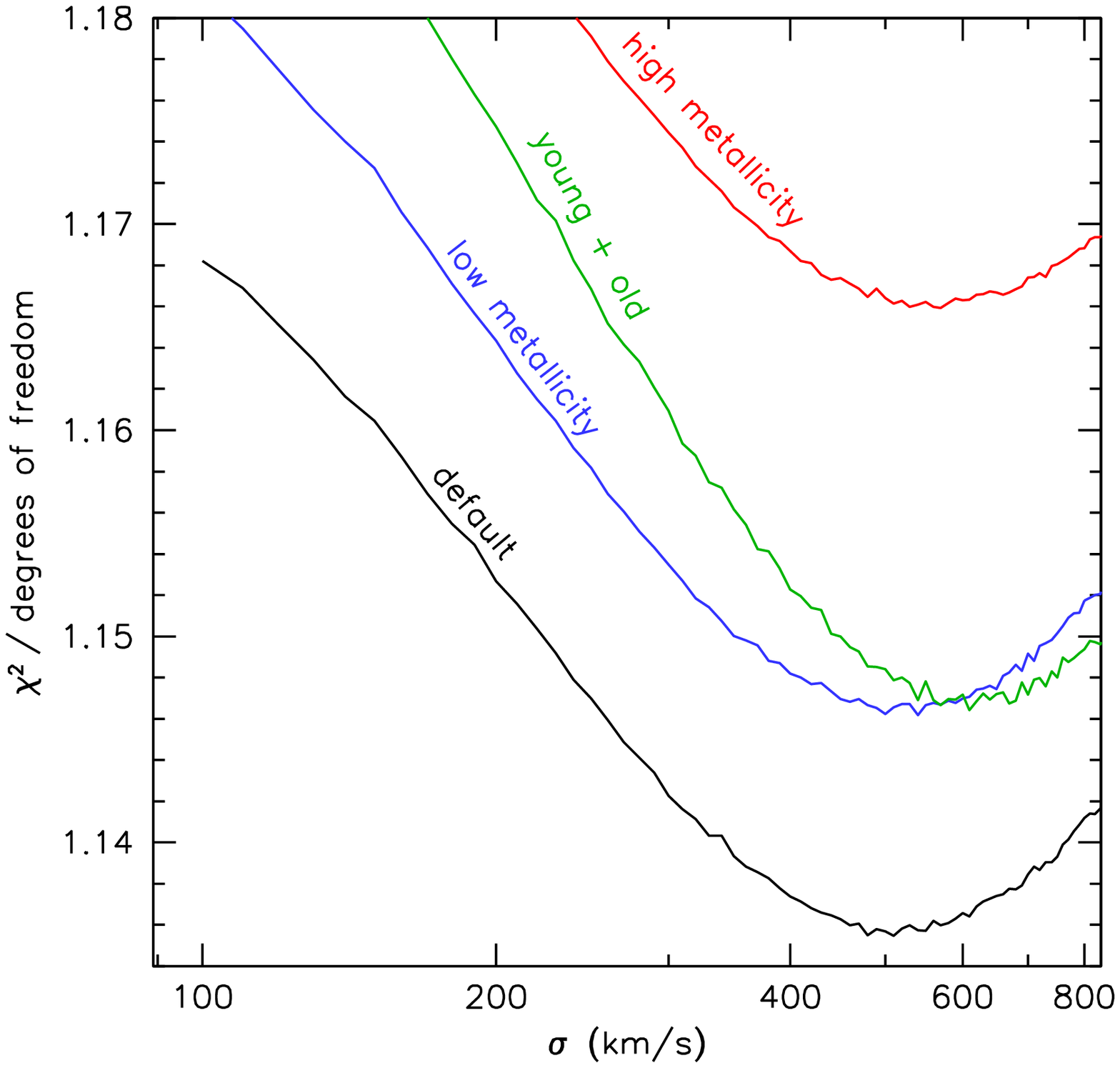}\vspace{0.3cm}\\
{\small \bf \textsf{Supplementary Figure 1} $|$ } 
{\small \bf \textsf{Reduced $\chi^2$ as a function of velocity dispersion.}}
{\small
The black curve shows results for the default template, which is the
best-fit stellar population synthesis model to the spectrum and
broad-band photometry. There is a well-defined minimum at
$\approx 500$\,\kms. The blue, red, and green curves are examples
of alternative templates, and illustrate the
effects of changing the template. The blue curve is for a single-age
stellar population with age 2\,Gyr and metallicity $0.4 \times$ Solar;
the red curve is for a 2\,Gyr population with metallicity
$2.5\times$ Solar;
and the green curve is for a composite model comprised
of a 0.5\,Gyr old component
in addition to a maximally-old compoment. These alternative
templates have higher
minimum $\chi^2$ values
than the default template, as expected. Importantly, the minima
occur at approximately the same dispersion as for the default template.}
\end{figure}

\noindent
\begin{figure}
\epsfxsize=13cm
\epsffile{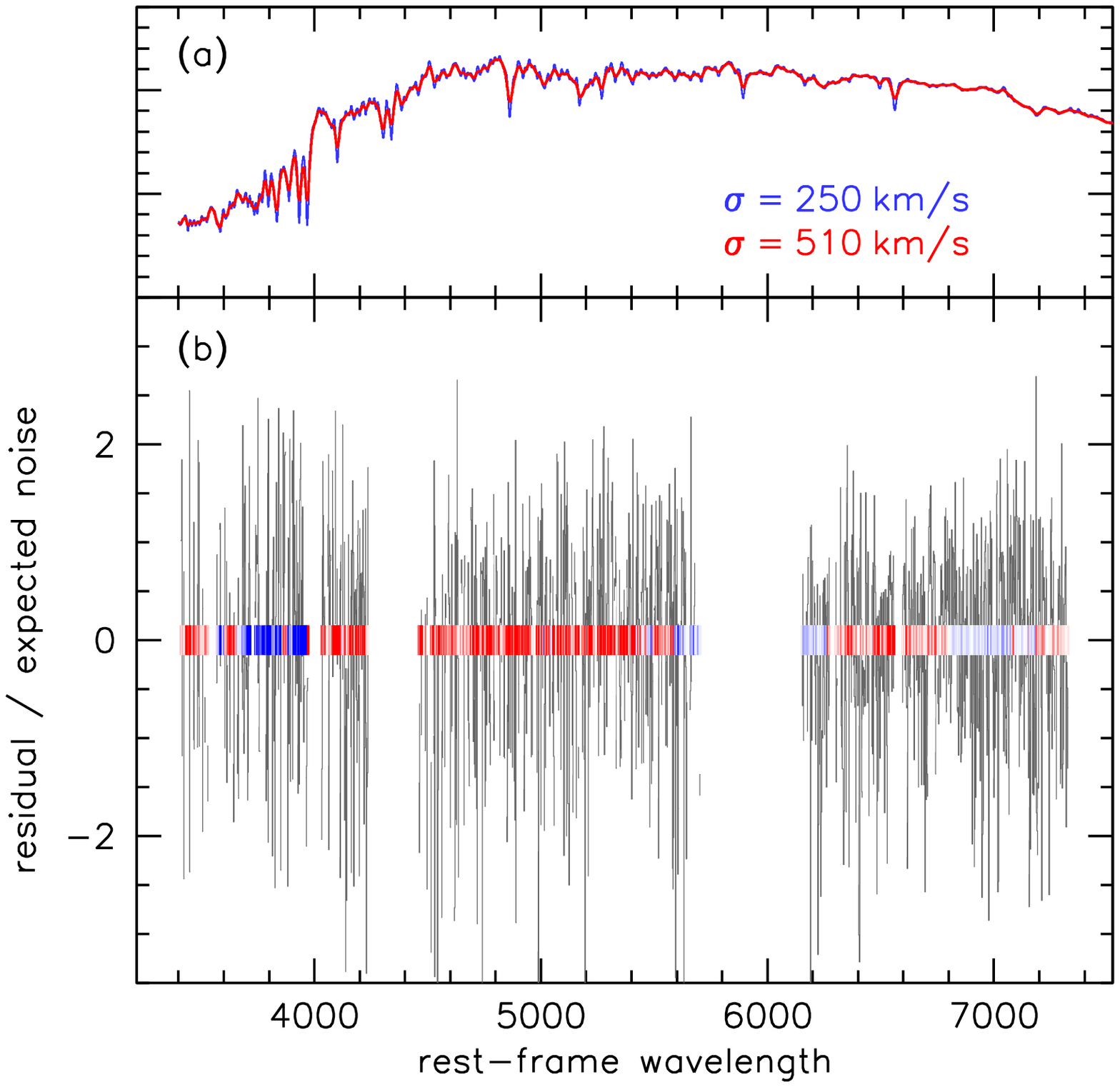}\vspace{0.3cm}\\
{\small \bf \textsf{Supplementary Figure 2} $|$ } 
{\small \bf \textsf{Analysis of residuals from the fit.}}
{\small {\bf a}, Models with
two different velocity dispersions. The red model has the best-fit
velocity dispersion of 510\,\kms\ and the blue model has a
dispersion that is a factor of two lower. The differences between
the models are subtle for any individual feature but are
significant when the entire spectrum is considered. {\bf b},
Residuals from the best-fit spectrum, normalized by the expected
noise, are shown in grey. The residuals are well-behaved, and the
reduced $\chi^2$ for the best-fit is close to 1. The colored band
shows which of the two models provides the best fit as a function
of wavelength. Each vertical bar is a median of the nearest 50 datapoints
(corresponding to $\approx 100$\,\AA\ in the rest-frame).
The hue (red or blue) indicates which model fits best and
the intensity indicates the absolute size of the
difference between the residual from
the low dispersion fit and the residual from the high dispersion
fit, with white implying that both fits are equally good.
With the exception of the region just blueward of the
4000\,\AA\ break the high-dispersion model provides a better
fit than the low-dispersion model. This demonstrates that our
results are not driven by a small wavelength region or by
other (obvious) wavelength-dependent effects.}
\end{figure}


\begin{thebibliography}{10}

\bibitem[{Trujillo} {\it et~al.}<1>]{trujillo:06}
{Trujillo}, I., {F\"orster Schreiber}, N.~M., Rudnick, G.,
Barden, M., Franx, M. {\it et al.}
{The Size Evolution of Galaxies
since $z\sim 3$: Combining SDSS, GEMS, and FIRES.}
\newblock {\it Astrophys.\ J.} {\bf 650}, 18--41 (2006).

\bibitem[{Toft} {\it et~al.}<2>]{toft:07}
{Toft}, S., {van Dokkum}, P., {Franx}, M., {Labb\'e}, I., {F\"orster
Schreiber}, N.~M. {\it et al.}
{Hubble Space Telescope and Spitzer Imaging of Red and Blue
Galaxies at $z\sim 2.5$: A Correlation between Size and Star
Formation Activity from Compact Quiescent Galaxies to Extended
Star-forming Galaxies.}
\newblock {\it Astrophys.\ J.} {\bf 671}, 285--302 (2007).

\bibitem[{van Dokkum} {\it et~al.}<3>]{vd:08}
{van Dokkum}, P.~G., {Franx}, M., {Kriek}, M., {Holden}, B.,
{Illingworth}, G.~D. {\it et al.}
{Confirmation of the Remarkable Compactness of Massive Quiescent
Galaxies at $z\sim 2.3$: Early-Type Galaxies Did not Form in a Simple
Monolithic Collapse.} 
\newblock {\it Astrophys.\ J.} {\bf 677}, L5-L8 (2008).

\bibitem[{Cimatti} {\it et~al.}<4>]{cimatti:08}
{Cimatti}, A., {Cassata}, P., {Pozzetti}, L., {Kurk}, J.,
{Mignoli}, M. {\it et al.}
{GMASS ultradeep spectroscopy of galaxies at $z\sim 2$. II.
Superdense passive galaxies: how did they form and evolve?}
\newblock {\it Astron.\ Astrophys.} {\bf 482}, 21-35 (2008).

\bibitem[{van der Wel} {\it et~al.}<5>]{wel:08}
{van der Wel}, A., {Holden}, B.~P., {Zirm}, A.~W., {Franx}, M., {Rettura},
A. {\it et al.} {Recent Structural Evolution of Early-Type Galaxies:
Size Growth from $z=1$ to $z=0$.}
\newblock {\it Astrophys.\ J.} {\bf 688}, 48-58 (2008).

\bibitem[{Franx} {\it et~al.}<6>]{franx:08}
{Franx}, M., {van Dokkum}, P.~G., {F\"orster Schreiber}, N.~M.,
{Wuyts}, S., {Labb\'e}, I., \& {Toft}, S.
{Structure and Star Formation in Galaxies out to $z=3$: Evidence
for Surface Density Dependent Evolution and Upsizing.}
\newblock {\it Astrophys.\ J.} {\bf 688}, 770-788 (2008).

\bibitem[{Damjanov} {\it et~al.}<7>]{dam:09}
{Damjanov}, I., {McCarthy}, P.~J., {Abraham}, R.~G., {Glazebrook},
K., {Yan}, H. {\it et al.} {Red Nuggets at $z\sim 1.5$:
Compact Passive Galaxies and the Formation of the Kormendy Relation.}
\newblock {\it Astrophys.\ J.}, in press (arXiv:0807.1744).

\bibitem[{Naab} {\it et~al.}<8>]{naab:07}
{Naab}, T., {Johansson}, P.~H., {Ostriker}, J.~P., \& {Efstathiou},
G. {Formation of Early-Type Galaxies from Cosmological
Initial Conditions.}
\newblock {\it Astrophys.\ J.} {\bf 658}, 710-720
(2008).

\bibitem[{Fan} {\it et~al.}<9>]{fan:08}
{Fan}, L., {Lapi}, A., {De Zotti}, G., \& {Danese}, L.
{The Dramatic Size Evolution of Elliptical Galaxies and the Quasar Feedback.}
\newblock {\it Astrophys.\ J.} {\bf 689}, L101-L104
(2008).

\bibitem[{Bezanson} {\it et~al.}<10>]{bezanson:09}
{Bezanson}, R., {van Dokkum}, P.~G., {Tal}, T., {Marchesini}, D.,
{Kriek}, M. {\it et al.} {The Relation Between Compact, Quiescent
High Redshift Galaxies and Massive Nearby Elliptical Galaxies:
Evidence for Hierarchical, Inside-Out Growth.}
\newblock {\it Astrophys.\ J.}, in press (arXiv:0903.2044).

\bibitem[{Naab} {\it et~al.}<11>]{naab:09}
{Naab}, T., {Johansson}, P.~H., {Ostriker}, J.~P.
{Minor Mergers and the Size Evolution of Elliptical Galaxies.}
\newblock {\it Astrophys.\ J.}, submitted (arXiv:0903.1636).

\bibitem[{van der Wel} {\it et~al.}<12>]{wel:09}
{van der Wel}, A., {Bell}, E.~F., {van den Bosch}, F.~C.,
{Gallazzi}, A., \& Rix, H.-W. {On the Size and Co-Moving
Mass Density Evolution of Early-Type Galaxies.}
\newblock {\it Astrophys.\ J.}, submitted (arXiv:0903.4857).

\bibitem[{Hopkins} {\it et~al.}<13>]{hopkins:09}
{Hopkins}, P.~F., {Bundy}, K., {Murray}, N., {Quataert}, E.,
{Lauer}, T., \& {Ma}, C.-P. {Compact High-Redshift Galaxies are
the Core of Present-Day Massive Spheroids.}
\newblock {\it Astrophys.\ J.}, submitted (arXiv:0903.2479).


\bibitem[{Kriek} {\it et~al.}<14>]{kriek:09}
{Kriek}, M., {van Dokkum}, P.~G., {Labb\'e}, I., {Franx}, M.,
{Illingworth}, G.~D., Marchesini, D., \& Quadri, R.~F.
{An Ultra-Deep Near-Infrared Spectrum of a Compact
Quiescent Galaxy at $z=2.2$.}
\newblock {\it Astrophys.\ J.}, in press (arXiv:0905:1692).

\bibitem[{Kriek} {\it et~al.}<15>]{kriek:06}
{Kriek}, M., {van Dokkum}, P.~G., {Franx}, M., {Quadri}, R.,
{Gawiser}, E. {\it et al.} {Spectroscopic Identification
of Massive Galaxies at $z\sim 2.3$ with Strongly Suppressed
Star Formation.}
\newblock {\it Astrophys.\ J.} {\bf 649}, L71-L74 (2006).

\bibitem[{Kriek} {\it et~al.}<16>]{kriek:08}
{Kriek}, M., {van Dokkum}, P.~G., {Franx}, M., {Illingworth}, G.~D.,
{Marchesini}, D. {\it et al.} {A Near-Infrared Spectroscopic
Survey of $K$-selected Galaxies at $z\sim 2.3$: Redshifts and
Implications for Broadband Photometric Studies.}
\newblock {\it Astrophys.\ J.} {\bf 677}, 219-237 (2008).

\bibitem[{Franx} {\it et~al.}<17>]{fih:89}
{Franx}, M., \& {Illingworth}, G., \& {Heckman}, T.
{Major and minor axis kinematics of 22 ellipticals.}
\newblock {\it Astrophys.\ J.}, {\bf 344}, 613-636 (1989).

\bibitem[{van Dokkum} \& {Stanford}<18>]{vd:03}
{van Dokkum}, P.~G., \& {Stanford}, S.~A.
{The Fundamental Plane at $z=1.27$: First Calibration of the Mass
Scale of Red Galaxies at Redshifts $z>1$.}
\newblock {\it Astrophys.\ J.}, {\bf 585}, 78-89 (2003).

\bibitem[{Bernardi} {\it et~al.}<19>]{bernardi:06}
{Bernardi}, M., {Sheth}, R.~K., {Nichol}, R.~C, {Miller}, C.~J.,
{Schlegel}, D.\ {\em et al.} {A Search for the Most Massive
Galaxies: Double Trouble?}
\newblock {\it Astron.\ J.},
{\bf 391}, 1191-1199 (2006).

\bibitem[{Bernardi} {\it et~al.}<20>]{bernardi:08}
{Bernardi}, M., {Hyde}, J.~B., {Fritz}, A., {Sheth}, R.~K.,
{Gebhardt}, K., \& {Nichol}, R.~C. {A Search for the Most Massive
Galaxies -- II. Structure, Environment, and Formation.}
\newblock {\it Mon.\ Not.\ R.\ Astron.\ Soc.},
{\bf 391}, 1191-1199 (2008).

\bibitem[{Cenarro} \& {Trujillo}<21>]{cenarro:09}
{Cenarro}, A., \& {Trujillo}, I. {Mild Velocity Dispersion Evolution
of Spheroid-like Massive Galaxies since $z\sim 2$.}
\newblock {\it Astrophys.\ J.}, {\bf 696}, L43-46 (2009).

\bibitem[{Dekel} {\it et~al.}<22>]{dekel:09}
{Dekel}, A., {Birnboim}, Y., {Engel}, G., {Freundlich}, J.,
{Goerdt}, T., {\it et al.} {Cold streams in early massive hot
haloes as the main mode of galaxy formation.}
\newblock {\it Nature} {\bf 457}, 451-454 (2009).

\bibitem[{Greve} {\it et~al.}<23>]{greve:05}
{Greve}, T.~R., {Bertoldi}, F., {Smail}, I., {Neri}, R., {Chapman}, S.~C.,
{\it et al.} {An interferometric CO survey of luminous
submillimetre galaxies.}
\newblock {\it Mon.\ Not.\ R.\ Astron.\ Soc.}
{\bf 359}, 1165-1183 (2005).

\bibitem[{Bertoldi} {\it et~al.}<24>]{bertoldi:03}
{Bertoldi}, F., {Cox}, P., {Neri}, R., {Carilli}, C.~L.,
Walter, F., {\it et al.}
{High-excitation CO in a quasar host galaxy at $z=6.42$.}
\newblock {\it Astron.\ Astrophys.} {\bf 409}, L47-L50 (2003).

\bibitem[{Narayanan} {\it et~al.}<25>]{narayanan:08}
{Narayanan}, D., {Li}, Y., {Cox}, T.~J., {Hernquist}, L.,
{Hopkins}, P., {\it et al.}
{The Nature of CO Emission from $z\sim 6$ Quasars.}
\newblock {\it Astrophys.\ J.\ Supp.} {\bf 174}, 13-30 (2008).

\bibitem[{Boylan-Kolchin} {\it et~al.}<26>]{boylan:06}
{Boylan-Kolchin}, M., {Ma}, C.-P., {Quataert}, E.\
{Red mergers and the assembly of massive elliptical galaxies:
the fundamental plane and its projections.}
\newblock {\it Mon.\ Not.\ R.\ Astron.\ Soc.} {\bf 369},
1081-1089 (2006).

\bibitem[{Ferrarese} \& {Merritt}<27>]{ferrarese:00}
{Ferrarese}, L., {Merritt}, D.\
{A Fundamental Relation between Supermassive Black Holes and Their
Host Galaxies.}
\newblock {\it Astrophys.\ J.} {\bf 539}, L9-L12 (2000).

\bibitem[{Gebhardt} {\it et~al.}<28>]{gebhardt:00}
{Gebhardt}, K., {Bender}, R., {Bower}, G., {Dressler}, A.,
{Faber}, S.~M., {\it et al.} {A Relationship between
Nuclear Black Hole Mass and Galaxy Velocity Dispersion.}
\newblock {\it Astrophys.\ J.} {\bf 539}, L13-L16 (2000).


\end{thebibliography}
\end{document}